\documentstyle[12pt]{article}
\newcommand{\eeq}{\end{equation}}

\begin{document}
\renewcommand{\theequation}{\thesection.\arabic{equation}}
\vskip 2cm
\title{Runaway Collapse of Witten Vortex Loops}
\author{\\
Arne L. Larsen${}^{*} {}^{1}$\, and \,
Minos Axenides${}^{\dag} {}^{2}$}
\maketitle
\noindent
$^{1}${\em
Theoretical Physics Institute, Department of Physics, \ University of
Alberta, Edmonton, Canada T6G 2J1}
\\ $^{2}${\em Department of Physics, University of Crete, P.O. Box
2208, 71003
Iraklion, Greece}
\begin{abstract}
\hspace*{-6mm}We consider general properties of charged circular
cosmic strings
in a general family of world-sheet string models. We then specialize
to a model
recently proposed by Carter and Peter. This model was shown to give a
good
description of the features of the superconducting cosmic strings,
originally
discovered by Witten in a $U(1)\times {U}(\tilde{1})$ field-theory.
We derive
an explicit expression for the potential determining the dynamics of
the string
and we present explicit expressions for the string tension and energy
density,
as a function of string-loop radius, in the locally preferred rest
frame. We
also obtain explicit expressions for the wiggle and woggle speeds
(speeds of
transverse and longitudinal perturbations, respectively). We show
that the
contraction of the uniformly charged string is essentially governed
by the
string tension (for large loop radius) and a {\it finite} Coulomb
barrier (for
small loop radius).  We argue for
the unobstructed contraction of a uniformly charged loop over the  
Coulombic
barrier and its eventual collapse to a charged point. The implication
of such an effect to the possible formation of naked singularities,  
in
violation of the cosmic censorship hypothesis, is finally discussed.
\end{abstract}
\noindent
PACS numbers: 0420, 1110, 1117\\
\\
\noindent
$^{*}$Electronic address: alarsen@phys.ualberta.ca\\
$^{\dag}$Electronic address: axenides@talos.cc.uch.gr \newpage
\section{Introduction}
\setcounter{equation}{0}
Superconducting cosmic strings have attracted a lot of interest since
the
pioneering work by Witten \cite{wit}. Although charge-current
carrying
superconducting strings had been considered before \cite{nil} (see
also
\cite{bal}), the underlying spacetime field-theoretic explanation and
understanding was missing. In a subsequent
paper
\cite{wit1}, it was proposed that static circular
string-loops exist whose equilibrium configuration is characterized  
by an
exact balance between the string tension and the current induced  
magnetic
force (see also \cite{ole} for a discussion of static
string-loops based
on external magnetic fields). This idea initiated a series of papers
on static
string-loops [6-13], reaching a wide variety of conclusions.
They nevertheless shared the same opinion that from a purely
"mechanical"
point of view, it is possible for the current to balance out the
string
tension; the only question was whether the current necessary for the
balance
would be below or above the maximum current $J_{max}$ (the current at
which the
current-carrying ability saturates \cite{wit}) in realistic physical
situations. Apparently the issue was settled in \cite{dav1},
concluding that
the current necessary for  balancing the string tension is indeed
larger than
$J_{max},$ that is, the current can not ensure the existence of
static
string-loops. However, in \cite{dav1,dav2} it
was also proposed that a rotation of the
superconducting  string-loop might do the job of balancing the string
tension
(this is obviously impossible for an ordinary non-conducting
Nambu-Goto
string-loop); the resulting equilibrium configuration is thus
strictly speaking
stationary but not static. These results and ideas have later been
confirmed
and developed in a series of papers by Carter and Peter [14-19],
for a review see Ref. \cite{car} (chapter IV).

More or less independently, although inspired by these results, there
has been
some interest in the purely "mechanical" properties of non-rotating
charge-current carrying superconducting string-loops. In these
studies, the
charge-current carrying superconducting string is described directly
in terms
of a two-dimensional world-sheet action, argued (or believed) to
represent the
effective action describing the dynamics of the core of a relatively
thin,
smooth and slowly varying 4-dimensional field-theoretic cosmic
string, of the
kind described by Witten \cite{wit}. Physical phenomena such as
gravitational
and electromagnetic radiation, the existence of a maximal
superconducter
current, backreaction effects on the background etc, are usually
neglected in
these papers.

The most popular of such world-sheet models is given by \cite{vil}:
\begin{equation}
{\cal S}_I=-\int d\tau d\sigma\sqrt{-G}\left(
\mu+\frac{1}{2}G^{\alpha\beta}
(\Phi_{,\alpha}+A_\mu X^\mu_{,\alpha})
(\Phi_{,\beta}+A_\mu X^\mu_{,\beta})  \right),
\end{equation}
where $\mu$ is the string tension, $A_{\mu}$ is the external
electromagnetic
potential, $\Phi$ is a scalar field on the world-sheet and
$G_{\alpha\beta}$ is
the induced metric on the world-sheet:
\begin{equation}
G_{\alpha\beta}=g_{\mu\nu}X^\mu_{,\alpha} X^\nu_{,\beta}.
\end{equation}
This model was used in most of the early papers on superconducting
strings
[6-8, 10-11, 21].
The purely mechanical properties of non-rotating string-loops
described by the
model (1.1), were investigated in [22-26]. Basically the result (in a
flat
Minkowski background devoid of external electromagnetic potentials)
is that
static circular string-loops can be obtained provided the string
carries either
a uniform charge density or a current (or both).

Another world-sheet model which has been discussed in detail in the
literature
is given by \cite{nil}:
\begin{equation}
{\cal S}_{II}=-\mu\int d\tau d\sigma\sqrt{-\left(
1+\frac{G^{\alpha\beta}}{\mu}
(\Phi_{,\alpha}+A_\mu X^\mu_{,\alpha})
(\Phi_{,\beta}+A_\mu X^\mu_{,\beta})\right) G}
\end{equation}
The mechanical properties of non-rotating string-loops in this model
([27-30]
as well as [22, 23, 25]) are somewhat different from the properties
of the
string-loops in the model (1.1). In particular, static string-loops
can only be
obtained provided the string carries {\it both} charge density {\it
and}
current [27]. It follows that a non-rotating string-loop with uniform
charge
density, but no current, has the somewhat unusual and unexpected (at
least to
us) property that it will contract and eventually collapse to a
(charged)
point, no matter how large the original charge density of the string
is. This
may have been part of the reason why most authors in the early years
preferred
the model (1.1), which seems to behave more like what is expected for
a charged
ring; namely, the contraction of a uniformly charged string-loop in
the model
(1.1) will stop before collapse due to an infinitely high Coulomb
barrier.

It was therefore somewhat surprising when the numerical
investigations
\cite{pet3} showed that in some respects, the model (1.3) gives a
better
description of Wittens field-theoretic cosmic string, than the model
(1.1).
Furthermore, a result of the numerical studies \cite{pet3} was the
recent
proposal \cite{carpet} of a new and mathematically somewhat more
complicated
model, which appearently gives a much better description of the
Witten cosmic
string than any of the models (1.1), (1.3).
\vskip 12pt
\hspace*{-6mm}The purpose of the present paper is to consider
uniformly charged
string-loops in this new world-sheet model by Carter and Peter
\cite{carpet}.

The paper is organized as follows:
In Section 2, we consider general properties of charged circular
strings in a
general family of world-sheet string models. We then specialize to
the model
recently proposed by Carter and Peter \cite{carpet}. We derive an
explicit
expression for the potential determining the dynamics of the string.
In Section
3, we consider some physical properties of the string. In particular,
we
present explicit expressions for the string tension and energy
density as a
function of string-loop radius in the locally preferred rest frame,
and we
obtain explicit expressions for the wiggle and woggle speeds (speeds
of
transverse and longitudinal perturbations, respectively). We show
that the
contraction of a uniformly charged string in this model is
essentially governed
by the string tension (for large loop radius) and a {\it finite}
Coulomb
barrier (for small loop radius).  We point out that a uniformly
charged
circular string can be above the {\it finite} Coulomb barrier and
then collapse
classically by simply contracting to a charged point. In Section 4,
we give our
conclusions and we present some speculations relating our results to
the
possible formation of naked singularities.

We use units where $ 4 \pi \varepsilon_{o} = \hbar = c = 1,$ such
that
$ e \approx 1/137$ and $ G \approx 2.1 \times 10^{15} (kg)^{-2}$.
In these
units, the string tension has dimension of mass-squared:
\begin{equation}
\mu \equiv m^{2} \approx 10^{-21} (kg)^{2},
\end{equation}
where $m$ is essentially the "Higgs" mass, and the numerical value is
obtained
for a GUT-string (see for instance \cite{vilenkin}).

\section{Charge-Current Carrying Strings}
\setcounter{equation}{0}
The starting point of our analysis will be a charge-current carrying
string
described by the following action \cite{brandon}:
\begin{equation}
{\cal S}\;=\;\int\;d\tau d\sigma\;{\cal L}\;\sqrt{-G},
\end{equation}
where $\cal L$ is the Lagrangian density (being just a constant for
the
ordinary
Nambu-Goto string). A very general family of string models that
introduces
electromagnetic self-interactions on the string as well as couplings
to the
external gravitational and electromagnetic potentials in a
reparametrization
invariant way, is obtained by letting the Lagrangian density $\cal L$
depend
on the world-sheet projection of the gauge covariant derivative of a
world-sheet
scalar field $ \Phi$ \cite{brandon}:
\begin{equation}
{\cal L} = {\cal L} (\omega) ,\;\;\;\;\;\;\;\;  \omega =
G^{\alpha\beta} (
\Phi_{,\alpha} +
A_{\mu} X^{\mu}_{,\alpha}) (\Phi_{,\beta} + A_{\mu}
X^{\mu}_{,\beta}).
\end{equation}
The spacetime energy-momentum tensor $T^{\mu\nu}$ and electromagnetic
current
$J^{\mu}$ are given by:
\begin{equation}
T^{\mu\nu} = \frac{2}{\sqrt{-g}} \frac{\delta {\cal S}}{\delta
g_{\mu\nu}},
\end{equation}
\begin{equation}
J^{\mu} = \frac{1}{\sqrt{-g}} \frac{\delta {\cal S}}{\delta A_{\mu}},
\end{equation}
so that the total mass-energy $ M$ and charge $Q$ of the string are:
\begin{equation}
 M  = \int d^{3}\vec{x}\;\sqrt{-g}\;\;T^{0}_{\;\;0} ,
\end{equation}
\begin{equation}
Q\;=\;\int d^{3}\vec{x}\;\sqrt{-g}\;\;J_{0}.
\end{equation}
Explicit expressions for $T^{\mu\nu}$ and $J^{\mu}$ are:
\begin{eqnarray}
\sqrt{-g}\;T^{\mu\nu}\;=\;\int\;d\tau
d\sigma\;\sqrt{-G}\;\hspace*{-2mm}&[&\hspace*{-3mm}
{\cal L}\;G^{\alpha\beta}X^{\mu}_{,\alpha}X^{\nu}_{,\beta} -
2 \frac{d {\cal L}}{d \omega} G^{\alpha\gamma} G^{\beta\delta}
X^{\mu}_{,\gamma}
X^{\nu}_{,\delta} ( \Phi_{,\alpha} + A_{\rho}
X^{\rho}_{,\alpha})\nonumber\\
\hspace*{-2mm}&(&\hspace*{-3mm}\Phi_{,\beta}+
A_{\sigma}X^{\sigma}_{,\beta} )] \delta (X - X (\tau, \sigma)),
\end{eqnarray}
and:
\begin{equation}
\sqrt{-g} J^{\mu}\;=\; 2 \int d\tau d\sigma \;\sqrt{-G}\;
\frac{d {\cal L}}{d \omega} G^{\alpha\beta} (\Phi_{,\alpha} + A_{\nu}
X^{\nu}_{,\alpha}) X^{\mu}_{,\beta} \delta(X - X(\tau,\sigma)).
\end{equation}

We will now specialize to circular strings in flat Minkowski space
and
zero
external electromagnetic potential:
\begin{equation}
g_{\mu\nu}\;=\;\eta_{\mu\nu} ,\;\;\;\;\;\;\;\;\;\;   A_{\mu}=0.
\end{equation}
The circular strings are parametrized by:
\begin{equation}
t\;=\;E \tau ,\;\;\;\;\;\;r\;=\;r(\tau) ,\;\;\;\;\;\;
\theta\;=\;\frac{\pi}{2}
, \;\;\;\;\;\;\phi\;=\;\sigma,
\end{equation}
where $E$ is a constant with dimension of length$^{-1}$  (=mass) and
$r(\tau)$
is to be determined from the equations
of motion. We further consider $\Phi$ depending only on the
world-sheet time
$\tau$:
\begin{equation}
\Phi\;\;=\;\;\Phi(\tau).
\end{equation}
This is motivated by the fact that we will consider uniformly
charged strings with zero current along the string. It turns out that
eq.($2.11$)
precisely describes such strings.

Let us now consider the equations of motion as obtained from
eqs.($2.1$)-($2.2$)
for the circular strings, eqs.($2.10$)-($2.11$). It is  
straightforward to show
that
these
equations lead to \cite{larsen}:
\begin{equation}
\dot{r}^{2}\;=\;E^{2}\;-\;r^{2}\;\left( {\cal L} +
\frac{\Omega ^{2}}{2
r^{2}(d{\cal L}/d\omega)}\right)^
{2},
\end{equation}
\begin{equation}
\dot{\Phi}\;=\;\frac{\Omega\sqrt{E^2 - \dot{r}^{2
}}}{2r(d{\cal L}/d\omega)},
\end{equation}
\begin{equation}
\omega\;=\;\frac{-\Omega^{2}}{4 r^{2}(d{\cal L}/d\omega)^{2}}\; ,
\end{equation}
where $\Omega$ is a dimensionless integration constant.
It follows that:
\begin{equation}
\sqrt{-g}\;J_{0}\;=\;E\Omega\;\int d\tau d\sigma \;\delta(X -
X(\tau,\sigma)),
\end{equation}
\begin{equation}
\sqrt{-g}\;T^{0}_{\;\;0} = E^{2} \int d\tau d\sigma \;\delta( X - X
(\tau,\sigma)).
\end{equation}
The charge and mass-energy of the string are then given by:
\begin{equation}
Q\; =\; E \Omega\; \int\; d\tau d\sigma \;\delta(t - t( \tau)) =
2\pi \Omega,
\end{equation}
\begin{equation}
M  =E^{2} \int \;d\tau d\sigma\; \delta (t - t(\tau)) =
 2 \pi E,
\end{equation}
so that ($E,$ $\Omega$) are the energy and charge densities of the
string,
respectively. Notice also that
all
these results have been obtained without specifying ${\cal
L}(\omega$) at
all !

Different Lagrangians ${\cal L}$ essentially correspond to different
ways of
introducing
charge (and current) on the string. The two most popular models  
studied in the
literature
are (c.f. eqs.(1.1), (1.3)):
\begin{equation}
{\cal L}_{I} = - ( \mu + \frac{\omega}{2}),
\end{equation}
\begin{equation}
{\cal L}_{II} =- \mu \sqrt{ 1 + \frac{\omega}{\mu}}\;.
\end{equation}
For these models one can easily solve
eq.$(2.14)$ for $\omega=\omega(\Omega^{2}, r^{2}, \mu)$ and then
eq.$(2.12)$
becomes:
\begin{equation}
\dot{r}^{2} + V(r) = E^{2},
\end{equation}
where:
\begin{equation}
V_{I} = ( \mu r + \frac{\Omega^{2}}{2r})^{2},
\end{equation}
\begin{equation}
V_{II}=\mu ( \mu r^{2} + \Omega^{2}),
\end{equation}
determining the dynamics of the string.
These two models have been discussed extensively in the literature
[22-30], so
we shall
not repeat the results here. Furthermore, it was recently shown by
Carter and
Peter \cite{carpet}
that a much more realistic model is provided by:
\begin{equation}
{\cal L}_{III}\;=\;-\;\mu\;-\frac{\omega}{2} ( 1 + \frac{\omega}
{m_{*}^{2}})^{-1}\;,
\end{equation}
which, besides the string tension $ \mu=m^{2}$, depends on an
additional
parameter $ m_{*}^{2}$. It was shown \cite{carpet} that the model of
eq.$(2.24)$
gives a much better description of the original Witten vortex
\cite{wit}, than
do the
previously presented models of eqs.$(2.19)-(2.20)$.
In the present paper we will not address to this issue any further.
We
will
merely take the model of eq.$(2.24)$ as it stands and consider the
dynamics
 of
circular strings as derived from it.

In the case of ${\cal L}_{III},$ eq.($2.14$) becomes:
\begin{equation}
\Omega^{2} ( 1 + \frac{\omega}{m_{*}^{2}})^{4} + \omega r^{2} = 0,
\end{equation}
to be solved for $ \omega = \omega( r^{2}, \Omega^{2}, m_{*}^{2})$.
Introduce
the two dimensionless quantities:
\begin{equation}
x \equiv  | \frac{ m_{*}}{\Omega} | r \geq 0 , \;\;\;\;\;\;\;\;v
\equiv -
\frac{\omega}{m_{*}
^{2}}.
\end{equation}
Then eq.$(2.25)$ has two real solutions given by:
\begin{equation}
v_{\pm} = 1 + \frac{1}{2}\sqrt{W(x)} \pm
\frac{1}{2}\sqrt{-W
(x) +
\frac{2x^{2}}{\sqrt{W(x)}}}\;,
\end{equation}
where:
\begin{eqnarray}
W(x)& \equiv& \frac{Z(x)}{54^{1/3}}-
\frac{128^{1/3} x^{2}}{Z(x)}\;,
\end{eqnarray}
\begin{eqnarray}
Z(x)& \equiv & \left( 27 x^{4} +
3 \sqrt{3}
\sqrt{x^{6} ( 256 + 27 x^{4})}\;
\right)^\frac{1}{3}.\end{eqnarray}
The potential in eq.$(2.21)$ now takes the form (in dimensionless
quantities):
\begin{equation}
\frac{1}{\Omega^{2}m_{*}^{2}} V_{\pm} = x^{2} \left(
\frac{\mu}{m_{*}^2} -
\frac{v_{\pm}}
{2}(1-v_\pm)^{-1} + \frac{1}{x^{2}}(1 - v_{\pm})^{2}\right)^{2}.
\end{equation}
This expression, once written out completely using eqs.(2.26)-(2.29),
gives the
potential explicitly and analytically as a function of the
string-loop radius,
but it is not very enlightening.
It is more useful to consider the two asymptotic  regions $x
\rightarrow 0\;\;
(r\rightarrow 0)$
and $ x\rightarrow \infty \;\;( r \rightarrow \infty) $.
For $x \rightarrow 0,$ eq.$(2.27)$ has the expansion:
\begin{equation}
v = \sum_{i=0}^{\infty} a_{i} x^{i/2},
\end{equation}
where:
\begin{eqnarray}
a_{i}=\sum_{j=1}^{i+1} \;\;\sum_{k=1}^{i+2-j}\;\;
\sum_{l=1}^{
i-j-k+3}
a_{4+i-j-k-l} \;a_{j}\;a_{k}\;a_{l},\;\;\;\;\;\;\;\;a_{0}=1.
\end{eqnarray}
That is to say,  $\;a_{1}=\pm 1,\;\;\; a_{2}=\frac{1}{4},\;\;\;
a_{3
}=\mp
\frac{1}{32}\;$
etc. It follows that:
\begin{equation}
v_{\pm} = 1 \pm \sqrt{x} + \frac{x}{4} \mp \frac{1}{32} x^{3/2}
+ {\cal O}(x^{2}).
\end{equation}
The potential $(2.30)$ then is:
\begin{equation}
\frac{1}{\Omega^{2}m_{*}^{2}} V_{\pm} = 1 \pm 2\sqrt{x} + {\cal
O}(x),
\end{equation}
for $x\rightarrow 0,$ i.e. :
\begin{equation}
V_{\pm}(0) = \Omega^{2} m_{*}^{2} ,\;\;\;\;\;\;\;\;  V'_{\pm}(0)=\pm
\infty,
\end{equation}
where prime denotes derivative with respect to $x.$

For $x\rightarrow \infty,$ we find instead:
\begin{eqnarray}
v_{+} &=& x^{\frac{2}{3}} + {\cal O}(1),
\end{eqnarray}
\begin{eqnarray}
v_{-} &=& x^{-2} + {\cal O}(x^{-4}),
\end{eqnarray}
and then the potential $(2.30)$ is :
\begin{equation}
\frac{1}{\Omega^{2}m_{*}^{2}} V_{+} = (
\frac{\mu}{m_{*}^{2}}+\frac{1}{2})^{
2} x^{2}
+ {\cal O}(x ^{2/3}),
\end{equation}
\begin{equation}
\frac{1}{\Omega^{2}m_{*}^{2}} V_{-} = ( \frac{\mu}{m_{*}^{2}})^{2}
x^{2} +
{\cal O}
(1),
\end{equation}
for $ x\rightarrow \infty ,$ i.e. :
\begin{equation}
 V_{\pm}(\infty) = \infty.
\end{equation}

Notice that the results obtained for the potential $V_{+}$ are
somewhat similar
(although there are small differences) to results obtained
for the
model  $(2.19$) of Nielsen [22, 23, 25, 27-30], while $(2.20)$ gives
completely
different results [22-26], especially for $r \rightarrow 0.$ The
results
obtained for the potential $V_{-}$
differs from the results of both models (2.19)-(2.20), especially for
$r
\rightarrow 0.$

For the model (2.24), we have now shown that in both cases ($V_\pm)$
the
potential goes to infinity at spatial infinity but it
 takes
a finite value at $x=0$ ($r=0$).
So there is no infinite Coulomb barrier for these charged strings.
This is
confirmed by numerical plots of $V_{\pm},$ Fig.1. It follows that a
uniformly
charged circular
string in the model (2.24) will collapse provided:
\begin{equation}
E^{2} \geq V_{\pm}(0)\;=\;\Omega^{2}m_{*}^{2}.
\end{equation}
That is,  if $(2.41)$ is fulfilled, a charged string-loop will
collapse
classically by simply contracting to a charged point.

In the next section we turn to the physical interpretation of the
above
results. Are they merely
artifacts of an unphysical model or do they actually describe real
physics ?
\section{The Physical Interpretation}
\setcounter{equation}{0}
In this section we will consider the physical interpretation of the
results
obtained in Section 2. In particular we address the question of
whether
they are trustworthy or whether they have been simply obtained
through  an
extrapolation of the model $(2.24)$ outside its range of validity.

In
the
notation
of Ref. \cite{carpet}, the Lagrangian $(2.24)$ reads:
\begin{equation}
\tilde{\Lambda}\;=\; -m^{2} + \frac{{ \tilde{\chi}}}{2} (1 -
\frac{ \tilde{\chi}}{m_{*}^{2}})^{-1}\;,
\end{equation}
that is:
\begin{equation}
 \omega=- {\tilde{\chi}} = -m_{*}^{2} v.
\end{equation}
Since $v$ is positive for the
circular strings, we are always in the "Electric" range
\cite{carpet}. The
energy density $U$ and string tension $T,$ in the locally preferred  
rest
frame,
are then given by \cite{carpet}:
\begin{equation}
U\;=\; m^{2} - \frac{{ \tilde{\chi}}}{2} (1 - \frac{{\tilde{\chi}}
}{m_{*}^{2}})
^{-1}
+ { \tilde{\chi}} (1 - \frac{{ \tilde{\chi}}}{m_{*}^{2}})^{-2}\;,
\end{equation}
\begin{equation}
T\;=\;m^{2} - \frac{{ \tilde{\chi}}}{2} (1 - \frac{{ \tilde{\chi}}}
{m_{*}^{2}})
^{-1}\;.
\end{equation}
The wiggle and woggle speeds (i.e., the speeds of transverse and
longitudinal perturbations, respectively) are \cite{carpet}:
\begin{equation}
c_{E}^{2}\;=\; \frac{T}{U}\;=\;1 - \frac{ \tilde{\chi}
(1 - \frac{
\tilde{\chi}}{m_{*}^{2}})^{-2}} {m^{2}-\frac{ \tilde{\chi}}{2}
(1-\frac{{
\tilde{\chi}}}
{m_{*}^{2}})^{-1} + { \tilde{\chi}} (1-\frac{{ \tilde{\chi}}}
{m_{*}^{2}})^{-2}}\;\;,
\end{equation}
\begin{equation}
c_{L}^{2} = - \frac{d T}{d U}\;=\;\frac{1-\frac{{ \tilde{\chi}}}
{m_{*}^{2}}}
{1+ \frac{3{ \tilde{\chi}}}{m_{*}^{2}}}\;\;.
\end{equation}
Notice that by using eqs.(2.26)-(2.29), $U,\;T,\;c_{E}$ and $c_{L}$
are given
explicitly and analytically as functions of the string-loop radius.
These expressions hold when ${\tilde{\chi}} >0$. However in order to
ensure
that $c_{E}^{2} >0$, $c_{L}^{2}>0$ as well as $ c_{L} < c_{E} < 1 $
(the
condition $ c_{L} < c_{E}$ must be fulfilled for the model ($2.24$)
to describe
properly
the Witten vortex \cite{carpet}), we must further have \cite{carpet}:
\begin{equation}
0 < \frac{{ \tilde{\chi}}}{m_{*}^{2}}\;<\;1\;-\;\frac{3
m_{*}^{2}}{2(2m^{2}+
m_{*}^{2})}\;\;.
\end{equation}
In our notation, eq.$(3.7)$ reads:
\begin{equation}
0 <  v_{\pm} <  1 -  \frac{3 m_{*}^{2}}{2(2\mu + m_{*}^{2})}\;\;.
\end{equation}
This condition is certainly not fulfilled in general for our
solutions
$(2.27)$. In
fact,
$ v_{+} \geq 1$ so that the corresponding string solution must be
neglected as
being unphysical, since it
actually has an imaginary woggle speed $c_{L}.$ On the other hand,
$v_{-}
 \leq 1$
and it is therefore possible that eq.$(3.8)$ gets fulfilled for the
corresponding string solution. In Fig.$2$
 we
plot $(c_{E}, c_{L})$ as a function of the radius of this
string-loop.
Clearly $ 0 \leq c_{L} < c_{E} \leq 1, $  except for $ x \rightarrow
0$.
It means that the circular string, which is determined by eq.(2.21)
with the
potential $V_{-}$ given by eq.(2.30) is supposed
to be a good approximation to a circular Witten vortex for any loop
radius
larger
than some $r_{0}$. Moreover, by increasing $ m^{2}/m_{*}^{2}$ we can
make this
$r_{0}$ arbitrarily small. Physically we actually expect $
m^{2}/m_{*}^{2}>>
1,$ since $m_{*}^2$ represents the charge-carrier mass while $m^2$
represents
the Higgs mass \cite{carpet}.

In conclusion, the circular  string solution corresponding to the
potential
$V_{-}$ can essentially be trusted
everywhere. Therefore, a uniformly charged circular string that
fulfills
eq.($2.41)$ collapses classically to a charged point, as there is no
Coulomb
barrier preventing the
collapse.
\section{Conclusion and Discussion}
\setcounter{equation}{0}
We have considered charged circular cosmic strings in a world-sheet
model
recently proposed by Carter and Peter \cite{carpet}.  We derived an
explicit
expression for the potential determining the dynamics of the string,
and we
presented explicit expressions for the string tension, the energy
density and
the wiggle and woggle speeds, as a function of the string-loop
radius.  We
showed that a uniformly charged circular string in this model can be
above the
{\it finite} Coulomb barrier and then collapse classically by simply
contracting to a charged point. This suggests that the end-result of
contraction will be a Reissner-Nordstrom black hole (the end-result
of
contraction of an uncharged circular string is a Schwarzschild black
hole).
However, we shall now argue for the possibility (at least in
principle) of the
formation of a naked singularity. After the collapse, we expect that
spacetime
is described by the  Reissner-Nordstrom line-element:
\begin{equation}
ds^2=-a(r)dt^2+\frac{dr^2}{a(r)}+r^2( d\theta^2+\sin^2\theta
d\phi^2),
\end{equation}
where in our units:
\begin{equation}
a(r)=1-\frac{2GM}{r}+\frac{GQ^2}{r^2}.
\end{equation}
This line-element describes a naked singularity if:
\begin{equation}
GM^2<Q^2,
\end{equation}
where $M=2\pi E$ and $Q=2\pi \Omega.$
For ordinary matter it is usually expected that if the initial
conditions are
such that eq.(4.3) holds, then the Coulomb barrier will prevent the
collapse
from taking place (cosmic censorship), thus the naked singularity
will not
form.  Interestingly enough, this does not seem to be the case for
the charged
cosmic string considered in this paper (which is supposed to be a
good
approximation to the original field-theoretic Witten-vortex). For the
charged
string to actually collapse, we have the restriction (2.41), but
eqs.(2.41),
(4.3) are actually fully consistent if the following relation holds:
\begin{equation}
G m_{*}^2<1.
\end{equation}
Now since $m_{*}^2<<m^2,$ we have that $G m_{*}^2<<1$ when using the
numerical
values for a GUT-string (1.4), so eq.(4.4) is trivially fulfilled.
This
suggests the possibility of formation of a naked singularity if the
initial
charged circular cosmic string is prepared such that eq.(4.3) holds.
It should
be stressed, however, that we have completely neglected such physical
effects
as gravitational and electromagnetic radiation, backreaction etc,
which might
change the above conclusions (for instance as in \cite{louis}). We
hope to
return to these questions elsewhere.
\vskip 12pt
\begin{centerline}
{\bf Acknowledgements}
\end{centerline}
\vskip 6pt
\hspace*{-6mm}The research of M.A. was partially supported by
Danmarks
Grund-Forsknings Fond through its support for the establishment of
the
Theoretical Astrophysics Center, while A.L.L. was supported by NSERC
(National
Sciences and Engineering Research Council of Canada). A.L.L. also
thanks B.
Carter for discussions on the results of Ref. [32], prior to its
publication.

\newpage
\begin{centerline}
{\bf Figure Captions}
\end{centerline}
\vskip 24pt
\hspace*{-6mm}Fig.1.
The potential (2.30) determining the dynamics of a charged
string-loop. Notice
that only $V_{-}$ shows a Coulomb barrier,
which furthermore is finite.
\vskip 12pt
\hspace*{-6mm}Fig.2.
The wiggle and woggle speeds (3.5)-(3.6), as a function of
string-loop radius,
for the string corresponding to the potential $V_{-}.$ Clearly $ 0
\leq c_{L} <
c_{E} \leq 1, $ except for $r\rightarrow 0.$
\end{document}